\documentclass[4apaper
   ,final            
  ]
  {aipproc}

\layoutstyle{6x9}

\begin{document}

\title{Newman-Penrose quantities as valuable tools 
in astrophysical relativity}

\author{Marco Bruni}{
  address={Institute of Cosmology and Gravitation, University of
Portsmouth,  Portsmouth PO1 2EG, UK.}
}

\author{Andrea Nerozzi}{
   address={Institute of Cosmology and Gravitation, University of
Portsmouth,  Portsmouth PO1 2EG, UK.}
 }

\author{Frances White}{
   address={Institute of Cosmology and Gravitation, University of
Portsmouth,  Portsmouth PO1 2EG, UK.}
}

\begin{abstract}
In this talk I will briefly outline work in progress in two different contexts in astrophysical relativity, i.e.\ the study of rotating star spacetimes and the problem of reliably extracting gravitational wave templates in numerical relativity. In both cases the use of Weyl scalars and curvature invariants helps to clarify important issues.
\end{abstract}

\maketitle


\section{Introduction}

 The  Weyl scalars - the components of the Weyl tensor over a null tetrad - were known and used in relativity before the introduction of the Newman-Penrose (NP) formalism, but within the latter they acquired a new relevance. Here I will summarise the use of Weyl scalars as tools in two different  contexts in astrophysical relativity. First I will briefly summarize recent work \cite{bertietal} aimed at assessing the validity of the Hartle-Thorne (HT) slow-rotation approximation for  describing stationary axisymmetric rotating Neutron Stars (NSs),  introducing work in progress \cite{FWt}
   to extend the analysis. In this context the Weyl scalars are used to construct invariant measures of the deviation of the exterior spacetime from Petrov type D, in view of a possible development of a Teukolsky-like perturbative formalism for rotating NSs.

   In the second part I will outline how the Weyl scalars may be used in numerical relativity in order to construct a wave extraction formalism for simulations dealing with spacetimes that will settle to a perturbed black hole (BH) at late times. In this case the Teukolsky BH perturbation formalism \cite{teukolsky} is in principle applicable, but it is difficult to extract a BH background spacetime, i.e.\ the gravitational mass and angular momentum, from a given simulation. Introducing the notion of a quasi-Kinnersly frame (also used in \cite{bertietal}), in \cite{AN1,AN2}  a method was proposed that bypasses this difficulty, by not requiring a background, and allows direct wave extraction. I will  present here work in progress \cite{ANt} where the method is directly applicable.

\begin{figure}
\label{Francesfig}
\begin{minipage}{.45\columnwidth}
 \includegraphics[height=.22\textheight]{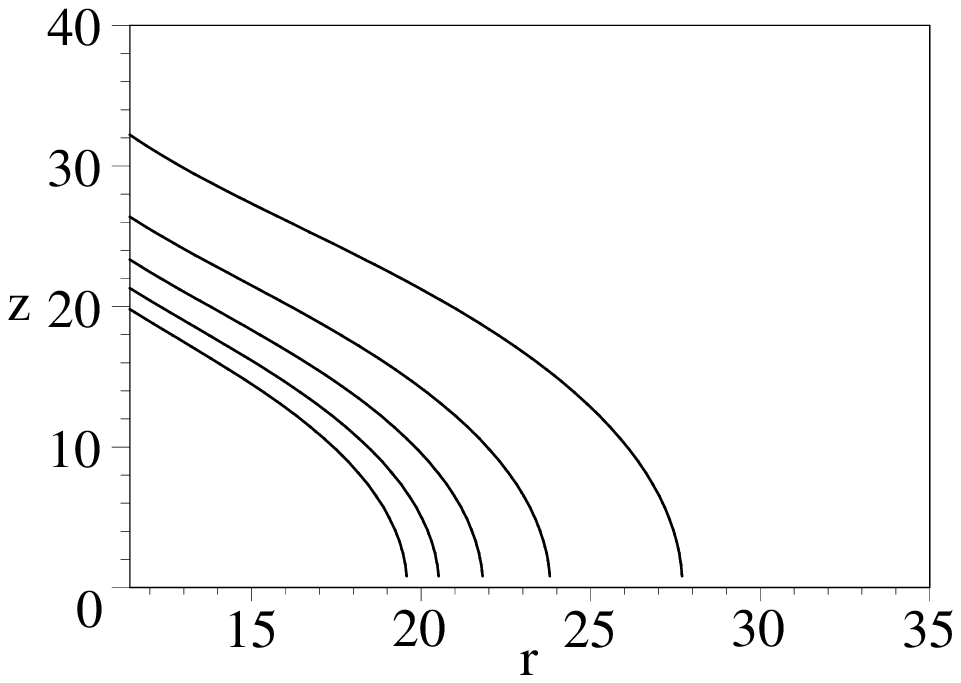}
   \end{minipage}
   \begin{minipage}{.44\columnwidth}
 \includegraphics[height=.22\textheight]{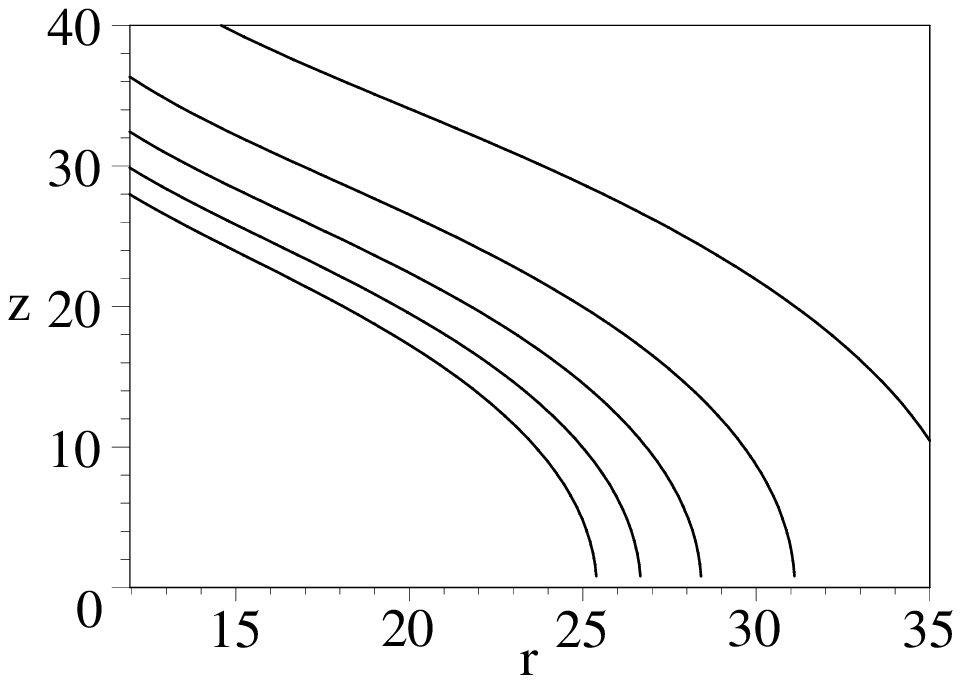}
   \end{minipage}
  \caption{Contour plots of $(1-|S|)\rm{x}10^{4}$ for decreasing values (from bottom left)  $5, 4, 3, 2, 1$, for a 1.4 $\rm{M}_\odot$ star, for a representative equation of state. Left panel: $\varepsilon=0.39258$; right panel: $\varepsilon=0.54440$.}
\end{figure}

\section{The spacetime of rotating stars}

Using a variety of models for NSs of different masses and equations of state, and comparing with full general relativistic numerical models, it was shown in \cite{bertietal} that the HT
 approximation to the metric of a rotating relativistic star  is very good for most astrophysical applications, even at the rotation rates of the fastest known milli-second pulsar.  For instance, the innermost stable circular orbit is predicted with an accuracy of $\sim 1\%$. It was also shown in \cite{bertietal} that although the spacetime of these stars is of Petrov type I (general; it would be type D for a spherical star),  the deviation from type D is small, at least on the equatorial plane. The HT metric is obtained as a perturbative solution  to second order in  $\epsilon=\Omega/\Omega^*$, where $\Omega$ is the star's angular
velocity and $\Omega^*=(M/R^3)^{1/2}$ is a "Keplerian" rotational scale.  The deviation from type D is measured by $1-|S|$, where $S=27J^2/I^3$ is an invariant curvature scalar, with $S=1$ for type D.
  Fig.\ \ref{Francesfig} \cite{FWt} shows that $1-|S|$ is small also out of the equatorial plane. In fact,  $1-|S|$ decreases more rapidly to zero out of the equatorial plane, as one would have expected. Thus in this sense $1-|S|$  in the equatorial plane is a good upper limit for the deviation from type D.

\section{Wave extraction}

\begin{figure}
\label{Andreafig}
  \includegraphics[height=.4\textheight]{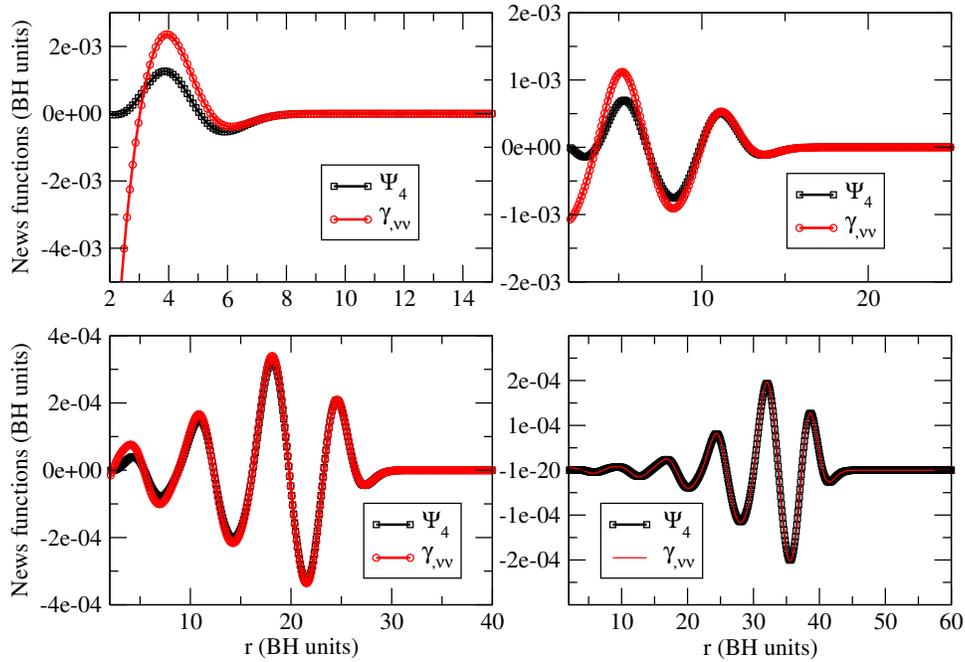}
  \caption{Comparison (from \cite{ANt}) of $\Psi_4$ and $\gamma_{,vv}$ at retarded times $v_1=1$, $v_2=20$, $v_3=50$, $v_4=80$.}
\end{figure}

 In \cite{AN1,AN2} methods were introduced to identify, for  a general numerical relativity implementation, what was dubbed the Quasi-Kinnersley frame, an equivalence class of tetrads that reduce, in the limit where the spacetime approaches type D, to the Kinnersly tetrad used in \cite{teukolsky} for the BH background. In this tetrad the Weyl scalar $\Psi_4$ carries information on outgoing gravitational radiation, and $\Psi_0$ on ingoing radiation. Work is in progress to identify one specific and physically significant tetrad from the equivalence class, appropriate for a generic numerical relativity code using an arbitrary ADM slicing. However, the method of \cite{AN1,AN2} is already applicable when using a null slicing, in particular to  the Bondi metric used in \cite{pap}, where non-linear oscillations of a BH  were analysed. In the case of the Bondi metric the gravitational wave signal can be extracted using the news function $\gamma_{,v}$ ($v$ is retarded time), directly related to the outgoing energy. In the linear regime, one expects $\Psi_4=-\gamma_{,vv}$. Thus this case is ideally suited to test our method, since we can compare  \cite{ANt} the news function obtained directly from the Bondi metric with that obtained  via $\Psi_4$ in the Quasi-Kinnersley frame. 
 Fig.\  \ref{Andreafig} shows the comparison of $\Psi_4$ with  $\gamma_{,vv}$, where  in our axisymmetric case $2\gamma=h^{TT}_{\theta\theta}$, i.e.\ the other polarization $h^{TT}_{\theta\phi}$ vanishes. Clearly the agreement is excellent at late times, as predicted, with an error $\Delta=|\Psi_4+\gamma_{vv}|\sim 10^{-6}$ at $v_4=80$.

\section{Conclusions} 

Here we have shown how the method obtained in \cite{AN1,AN2} works,   reproducing very well results obtained in \cite{pap}, when applied in the context of a code using the characteristic  approach \cite{ANt}. 
Work is in progress to find a  method to choose one particular Quasi-Kinnersly tetrad out of the general class in order to properly extract gravitational wave templates from any numerical relativity code using an arbitrary ADM slicing. 

For a rotating NS, we have shown that the deviation of the spacetime from Petrov type D is always very small, with  $1-|S|$ rapidly decreasing out of the equatorial plan and with increasing distance from the star.






\IfFileExists{\jobname.bbl}{}
 {\typeout{}
  \typeout{******************************************}
  \typeout{** Please run "bibtex \jobname" to optain}
  \typeout{** the bibliography and then re-run LaTeX}
  \typeout{** twice to fix the references!}
  \typeout{******************************************}
  \typeout{}
 }

\end{document}